\documentclass[aps,pra,reprint,superscriptaddress,longbibliography]{revtex4-1}

\usepackage{color}
\usepackage{graphicx}
\usepackage{epstopdf}
\usepackage{amsmath}
\usepackage{textcomp}
\usepackage{textgreek}
\usepackage{hyphenat}
\usepackage[hidelinks]{hyperref}
\usepackage[normalem]{ulem}

\begin{document}

\title{Higher-order meniscus oscillations driven by flow-focusing leading to bubble pinch off and entrainment in a piezo acoustic inkjet nozzle}

\author{Arjan Fraters}
\affiliation{Physics of Fluids group, Max-Planck Center Twente for Complex Fluid Dynamics, Department of Science and Technology, MESA+ Institute, and J. M. Burgers Centre for Fluid Dynamics, University of Twente, Enschede, Netherlands}
\author{Maaike Rump}
\affiliation{Physics of Fluids group, Max-Planck Center Twente for Complex Fluid Dynamics, Department of Science and Technology, MESA+ Institute, and J. M. Burgers Centre for Fluid Dynamics, University of Twente, Enschede, Netherlands}
\author{Roger Jeurissen}
\affiliation{Department of Applied Physics, Eindhoven University of Technology, Eindhoven, Netherlands}
\author{Marc van den Berg}
\affiliation{Canon Production Printing Netherlands B.V., Venlo, Netherlands}
\author{Youri de Loore}
\affiliation{Canon Production Printing Netherlands B.V., Venlo, Netherlands}
\author{Hans Reinten}
\affiliation{Canon Production Printing Netherlands B.V., Venlo, Netherlands}
\author{Herman Wijshoff}
\affiliation{Canon Production Printing Netherlands B.V., Venlo, Netherlands}
\affiliation{Department of Mechanical Engineering, Eindhoven University of Technology, Eindhoven, Netherlands}
\author{ Devaraj van der Meer}
\affiliation{Physics of Fluids group, Max-Planck Center Twente for Complex Fluid Dynamics, Department of Science and Technology, MESA+ Institute, and J. M. Burgers Centre for Fluid Dynamics, University of Twente, Enschede, Netherlands}
\author{Detlef Lohse}
\affiliation{Physics of Fluids group, Max-Planck Center Twente for Complex Fluid Dynamics, Department of Science and Technology, MESA+ Institute, and J. M. Burgers Centre for Fluid Dynamics, University of Twente, Enschede, Netherlands}
\author{Michel Versluis}\vspace{6pt}
\affiliation{Physics of Fluids group, Max-Planck Center Twente for Complex Fluid Dynamics, Department of Science and Technology, MESA+ Institute, and J. M. Burgers Centre for Fluid Dynamics, University of Twente, Enschede, Netherlands}
\author{{Tim Segers}}
\affiliation{BIOS Lab-on-a-Chip group, Max-Planck Center Twente for Complex Fluid Dynamics, MESA+ Institute for Nanotechnology, University of Twente, Enschede, Netherlands}
\affiliation{Physics of Fluids group, Max-Planck Center Twente for Complex Fluid Dynamics, Department of Science and Technology, MESA+ Institute, and J. M. Burgers Centre for Fluid Dynamics, University of Twente, Enschede, Netherlands}

\begin{abstract}
The stability of high-end piezo-acoustic drop-on-demand (DOD) inkjet printing is sometimes compromised by the entrainment of an air bubble inside the ink channel. Here, bubble pinch-off from an acoustically driven meniscus is studied in an optically transparent DOD printhead as a function of the driving waveform. We show that bubble pinch-off follows from low-amplitude higher-order meniscus oscillations on top of the global high-amplitude meniscus motion that drives droplet formation. In a certain window of control parameters, phase inversion between the low and high frequency components leads to the enclosure of an air cavity and bubble pinch-off.  Although phenomenologically similar, bubble pinch-off is not a result of capillary wave interaction such as observed in drop impact on a liquid pool. Instead, we reveal geometrical flow focusing as the mechanism through which at first, an outward jet is formed on the retracted concave meniscus. When the subsequent high-frequency pressure wave hits the now toroidal-shaped meniscus, it accelerates the toroidal ring outward resulting in the formation of an air cavity that can pinch off. The critical control parameters for pinch off are the pulse timing and amplitude. To cure the bubble entrainment problem, the threshold for bubble pinch-off can be increased by suppressing the high frequency acoustic waves through appropriate waveform design. The present work therefore aids the improvement of the stability of inkjet printers through a physical understanding of meniscus instabilities. 

\end{abstract}

\maketitle

\section{Introduction}

Piezo inkjet printing is an accurate and contactless method to deposit ink droplets on a substrate~\cite{Basaran2002,wijshoff2010inkjet, hoath2015fundamentals, Lohse2021}. Droplets are formed on-demand from a nozzle by actuating a piezoelectric element. The piezo deforms the channel wall upon electrical stimulation, resulting in acoustic pressure waves that jet the ink out of the nozzle~\cite{Fraters2020secondary}. Piezo inkjet printing is used in high-end industrial printers for on-demand personalized printing of documents, graphic art, and packaging. The main reason being its high reliability, high print quality, and its compatibility with a wide range of inks. The aforementioned properties make piezo inkjet printing also an excellent technique for several emerging additive manufacturing applications such as printing electronics~\cite{majee2016graphene, majee2017graphene, eshkalak2017cnt, vilardell2013ceramic, moya2017electrochemical, eggenhuisen2015solarcells, hashmi2015solarcells, shimoda2003displays, jiang2017displays}, pharmaceutics~\cite{daly2015pharmaceutics}, biomaterials \cite{simaite2016muscles, hewes2017microvessels, nakamura2005tissue, villar2013tissue}, and even for the lubrication of ball bearings~\cite{Kruk2019}.

\begin{figure*}[t]
\includegraphics[width=0.81\textwidth]{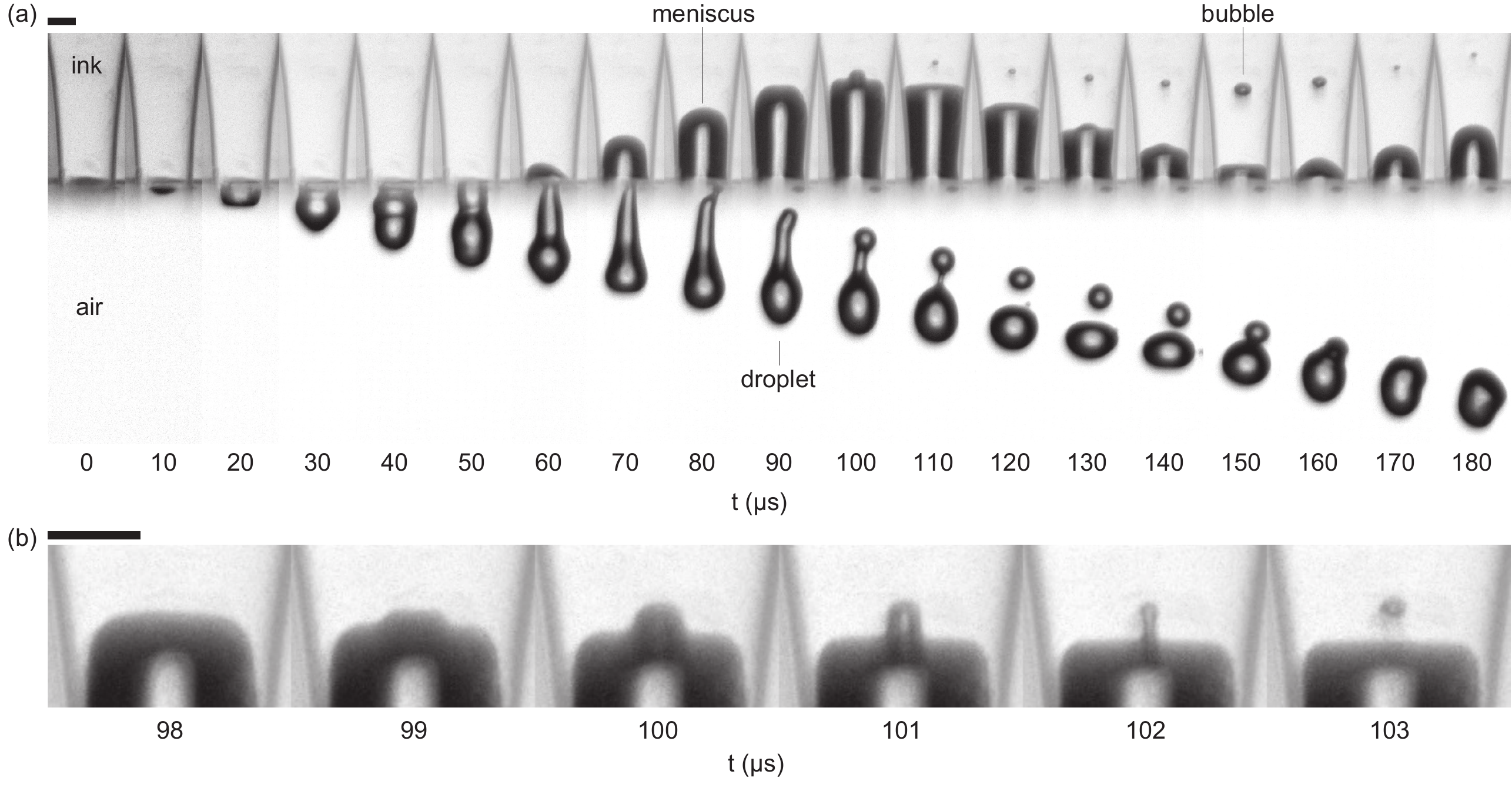}
\caption{(a) Bubble pinch-off and entrainment in a 70~\textmu m diameter nozzle of a piezo drop-on-demand inkjet printhead. The piezo actuation pulse was a rectangular push-pull pulse with a 150~V amplitude and a 30~\textmu s width. The images were recorded using 8~ns single-flash stroboscopic imaging with illumination by laser-induced fluorescence (iLIF)~\cite{vanderbos2011ilif}. (b) Details of the bubble pinch-off process: The center of the meniscus moves inward while the outer region of the meniscus moves outward, leading to the formation of an air cavity that eventually pinches off as an entrained air bubble. The scale bars represent 50~\textmu m.}\label{fig:1}
\end{figure*}

Although piezo inkjet printing is a highly reliable droplet deposition technique, the droplet formation process is sometimes compromised by the entrainment of an air bubble~\cite{dejong2006airentrapment, dejong2006entrapped, jeurissen2008effect, jeurissen2009acoustic, lee2009dynamics, kim2009effects, jeurissen2011regimes, vanderbos2011infrared,Fraters2019Infrared,Fraters2019Nucleation}. The entrained air bubble disturbs or even stops the jetting process and thereby dramatically reduces the printing quality and reliability. Previously, several mechanisms have been identified by which bubbles can be entrained in the ink channel. First, on the nozzle plate at the nozzle exit, dirt particles or an ink layer can trigger bubble entrainment by disturbing the jetting process at the nozzle exit~\cite{dejong2006airentrapment}. Second, a dirt particle in the ink can trigger bubble nucleation upon its interaction with the oscillating meniscus interface and, third, a bubble can nucleate on the particle through cavitation inception in the rarefaction pressure wave~\cite{Fraters2019Nucleation}.  However, bubbles can also be entrained in the absence of dirt particles or an ink layer, i.e.,~by yet another physical mechanism. Figure~\ref{fig:1} shows such a bubble pinch-off and entrainment event that was observed in a squeeze type piezo inkjet printhead with a 70-\textmu m diameter optically transparent nozzle exit (Microdrop Technologies GmbH, Autodrop Pipette AD-K-501), driven by a rectangular push-pull pulse (amplitude:~150~V, width:~30~\textmu s). First, a droplet is ejected, and subsequently, the meniscus retracts back into the nozzle and a bubble pinches off when the meniscus motion reverses from its inward motion to an outward motion, away from the ink channel. The bubble pinch-off event is shown in more detail in Fig.~\ref{fig:1}(b). The figure shows that the central region of the meniscus moves inward while the outer region of the meniscus moves outward. As a result, an air cavity forms that eventually closes, thereby pinching off an air bubble.

\begin{figure}[b]
\includegraphics[width=1\columnwidth]{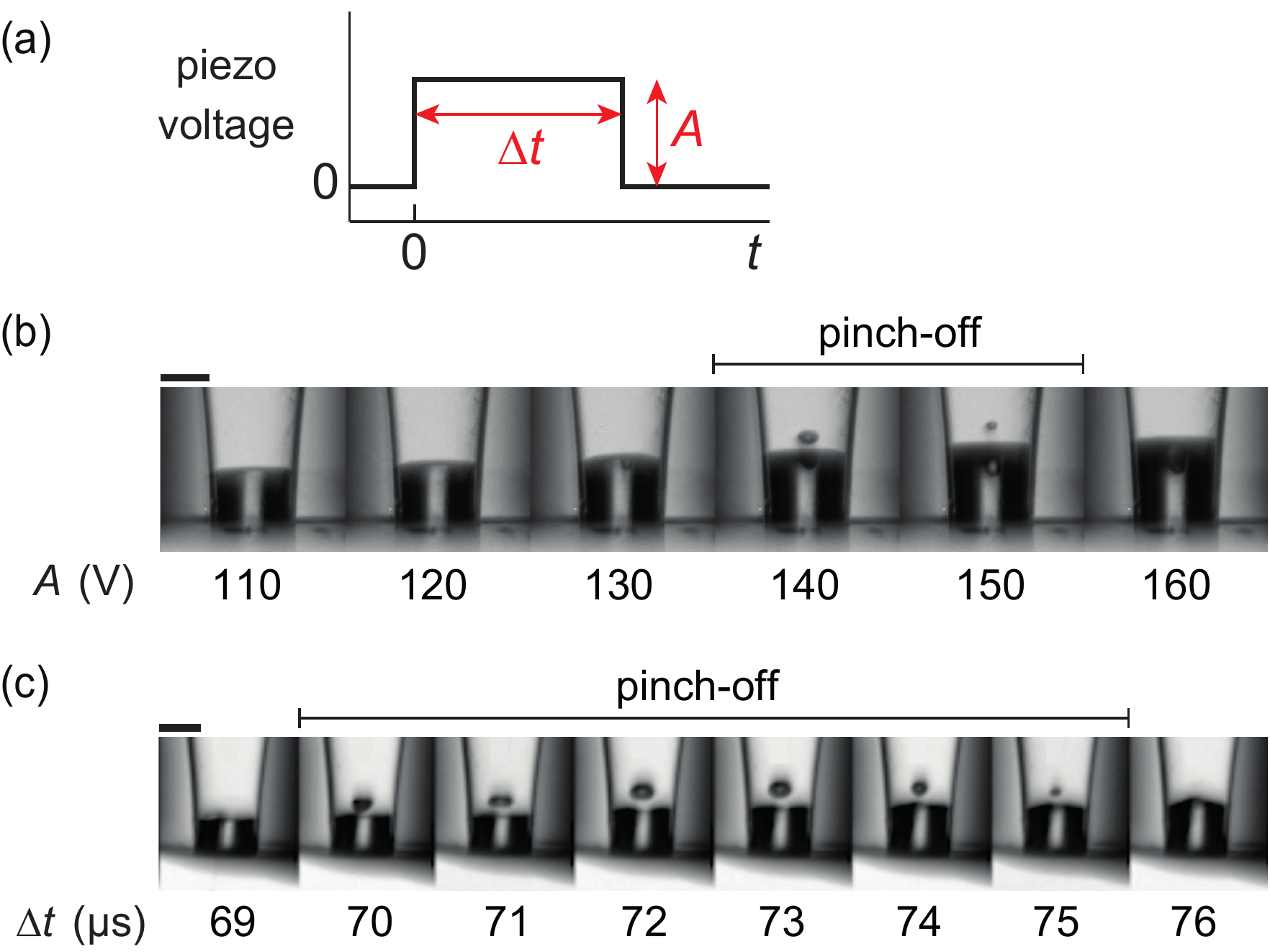}
\caption{(a) Rectangular piezo actuation pulse with its amplitude $A$ and length $\Delta t$ as the control parameters. Window of bubble pinch-off for (b) a pull-push pulse with a pulse width of 30~\textmu s and a varying amplitude, and (c) a pull-push pulse with an amplitude of 94~V and a varying pulse width. The scale bars represent 50~\textmu m.}\label{fig:2}
\end{figure}

Bubble pinch-off as shown in Fig.~\ref{fig:1} was found to occur only within certain windows of the piezo driving conditions. This is illustrated in Fig.~\ref{fig:2}, where two examples of a bubble pinch-off window are given for a rectangular pull-push pulse with amplitude $A$ and width $\Delta t$ (Fig.~\ref{fig:2}(a)). In the first example in Fig.~\ref{fig:2}(b), the pulse amplitude $A$ was varied with all other parameters fixed. A window of bubble pinch-off was observed between pulse amplitudes of 140~V and 150~V. Given the nature of meniscus instabilities, meaning that the growth time shortens and the oscillation amplitude increases with increasing acceleration~\cite{rayleigh1883investigation, taylor1950instability, faraday1831peculiar}, it was expected that bubble pinch-off would always occur above a certain threshold amplitude. Surprisingly, no bubble pinch-off was observed at amplitudes larger than 160~V. In the second example, see Fig.~\ref{fig:2}(c), the pulse width was varied. Bubble pinch-off was observed between pulse widths of 70~\textmu s and 75~\textmu s. The bubble size initially increases and then decreases, with a maximum radius between 72~\textmu s and 73~\textmu s. 

An oscillating meniscus can be destabilized by several mechanisms, including the classical Rayleigh-Taylor instability~\cite{rayleigh1883investigation, taylor1950instability,vdMeulen2020} and the parametrically driven meniscus instability~\cite{faraday1831peculiar, tence1997method}. A Rayleigh-Taylor instability grows on a flat interface between two fluids with different density, i.e., in this particular case the ink and air. The two fluids are accelerated at a rate high enough for the inertial forces to overcome the restoring surface tension. The parametrically driven meniscus instability grows on an initially flat meniscus at the subharmonic of the frequency at which the meniscus is driven (period doubling). The meniscus can also be destabilized at intermediate Ohnesorge number by an inhomogeneous velocity field at the meniscus due to the finite transport time of viscous-drag-induced vorticity from the wall to the center of the nozzle~\cite{chen2002anewmethod}. Furthermore, the meniscus can be deformed by geometrical flow focusing when a pressure wave hits a concave meniscus~\cite{antkowiak2007short-term,peters2013highly,Gordillo2020}. Finally, meniscus destabilization and resulting bubble pinch-off can originate from the interaction of capillary waves at the gas-liquid interface. In fact, there is a remarkable similarity between the bubble entrainment phenomenon in Figs.~\ref{fig:1} and~\ref{fig:2}, and in particular the window of bubble entrainment, with bubble pinch-off during crater collapse in drop impact and that during bubble bursting at the surface of a liquid pool~\cite{Oguz1990,Pumphrey1990,Thoroddsen2018,Sleutel2020,Duchemin2002,Gordillo2019}. 
In these cases, bubble entrainment is a result of capillary waves traveling down the cavity that then interact at the base of the cavity leading to the pinch-off of a bubble. Bubble pinch-off has been shown to require perfect timing of capillary wave interaction and it thereby depends on both the amplitude and the overal shape of the cavity~\cite{Oguz1990,Sleutel2020,Gordillo2019}. 

The aim of the present study is to find the underlying physical mechanisms that drive bubble pinch-off and entrainment as observed in Figs.~\ref{fig:1}~and~\ref{fig:2}, and to gain fundamental physical insight into the stability of an acoustically driven meniscus. To that end, after describing the methods for both the experiments and numerics in Section II, the meniscus and bubble dynamics shown in Fig.~\ref{fig:1} are analyzed in more detail by tracking the meniscus position over time (Section III). The acoustic driving of the meniscus by the piezo actuator is further characterized by measuring the ring-down piezosignal. Then the meniscus dynamics of two additional experiments is analyzed to identify the process during which the inner and and outer region of the meniscus develop their destabilizing out-of-phase motion. Finally, the mechanisms that drive the development of this out-of-phase motion are identified using numerical simulations with the boundary integral (BI) method~\cite{peters2013highly}. The paper ends with conclusions (section IV).

\section{Experimental and numerical methods}

\subsection{Printhead and ink}

\begin{figure}[b] 
\includegraphics[width=1\columnwidth]{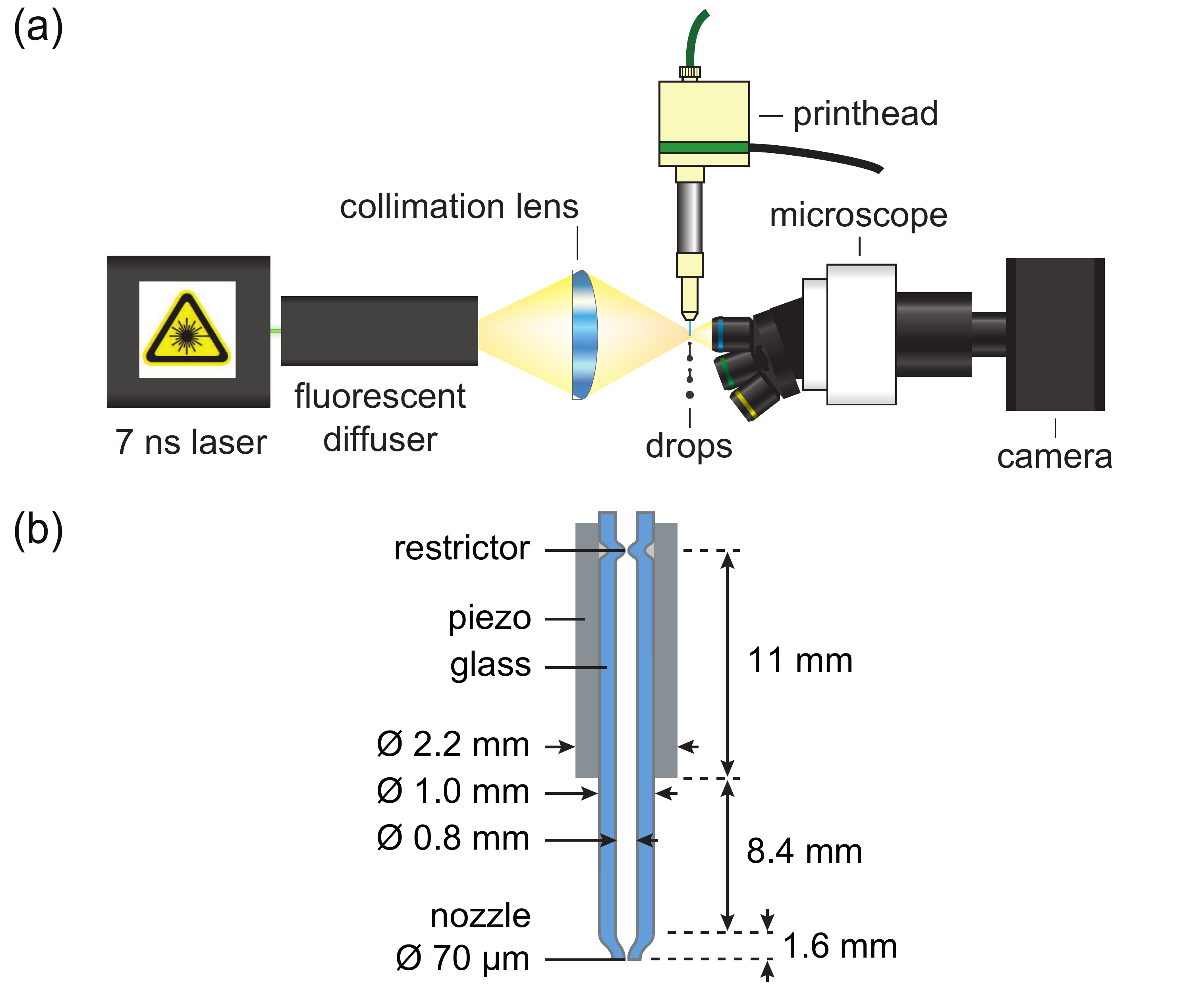}
\caption{(a) Experimental setup employed to image meniscus motion and bubble pinch-off in the drop-on-demand piezo-acoustic inkjet nozzle using illumination by laser-induced fluorescence (iLIF)~\cite{vanderbos2011ilif}. (b) Schematic layout of the functional acoustic part of the inkjet printhead. It consists of a glass capillary tube (blue) that is tapered towards the 70~\textmu m nozzle exit.  A cylindrical piezo (gray) can be actuated to drive the ink channel acoustics and the resulting droplet formation.}\label{fig:3}
\end{figure}

A 70~\textmu m nozzle diameter Autodrop Pipette from Microdrop Technologies GmbH (AD-K-501 and AD-H-501) was used, see Fig.~\ref{fig:3}(a). Figure~\ref{fig:3}(b) shows the approximate inner dimensions of the functional acoustic part of the printhead. More details about this type of printhead can be found in refs.~\cite{dijksman1984hydrodynamics, dijksman1998hydro-acoustics}.

A 4:1 (v/v) mixture of water with glycerol (Sigma-Aldrich, G9012, 1,2,3-Propanetriol, $\geq$99.5\%) was used as a model ink. All experiments were performed at room temperature. The density, viscosity, and surface tension $\sigma$ were taken from literature to be 1050~kg/m$^{3}$, 2.1~mPa$\cdot$s, and 71~mN/m, respectively~\cite{segur1951viscosity, glycerine1963physical}. The model ink was supplied from a plastic syringe to the top of the Autodrop Pipette holder via flexible plastic PEEK tubing (Upchurch Scientific), and the meniscus was positioned at the nozzle exit by manually adjusting the piston of the syringe.

\subsection{Imaging setup}

Bubble pinch-off was recorded using a stroboscopic imaging setup, see Fig.~\ref{fig:3}(a). The microscope (Olympus) had a 5$\times$ objective (LMPLFLN5x), a tube lens (U-TLU), and a high-resolution CCD camera (Lumenera, Lw135m, 1392$\times$1040~pixels, 4.65~\textmu m pixel size). The resulting optical resolution was 0.93~\textmu m/pixel. The images captured by the camera were saved by custom-made software on a Personal Computer (PC) programmed in the graphical programming language Labview (National Instruments).

The tip of the Autodrop Pipette was illuminated by incoherent 8~ns illumination pulses with a wavelength of 577~nm from a Laser-Induced Fluorescence (iLIF) system \cite{vanderbos2011ilif}. The iLIF system consisted of a pulsed laser (Quantel EverGreen, dual cavity Nd:YAG, \textlambda = 532~nm, 7~ns), a fluorescent plate embedded in a highly efficient diffuser (Lavision, part nr. 1108417 and 1003144), and a lens to condense the light pulses onto the imaging plane of the microscope.

\subsection{Measurement procedure}

A programmable pulse-delay generator (Berkeley Nucleonics Corp., BNC 575) triggered the laser, the camera, and the printhead actuation system with nanosecond precision. The jetting process was kept reproducible by jetting the entrained bubble outwards after each bubble pinch-off event. To do so, the piezo was actuated by rectangular pulses from two arbitrary waveform generators: one waveform generator (Agilent 33220A, 20~MHz, 14~bit, 50~MS/s) produced one high-amplitude pulse to entrain an air bubble, and the other waveform generator (Wavetek 195, 16~MHz, 12~bit, 40~MS/s) produced successively 49 low amplitude pulses to jet the entrained air bubbles out of the nozzle. For every actuation cycle, a custom-made Labview program captured one image during the high-amplitude piezo actuation pulse. The timing of image exposure was controlled by varying the delay of the laser flash with respect to the start of the piezo driving pulse. The delay was varied over a range from 0~\textmu s to 200~\textmu s with steps of 1~\textmu s to capture the complete drop formation and bubble pinch-off process. A laboratory amplifier (Falco System WMA-300, 5~MHz, 2000~V/\textmu s) amplified the pulses from the waveform generators by a factor of 50. Given the 5~MHz amplifier bandwidth, the rise- and fall time of the rectangular pulses was 0.2~\textmu s. The rectangular piezo driving pulses had an amplitude between 0~V and 160~V, and the printhead could be driven in either the push-pull mode or pull-push mode by switching the polarity of the electrical connections at the printhead. With the complete system, droplets were produced with diameters in the range of 70~\textmu m to 100~\textmu m, corresponding to volumes of 180~pL to 520~pL, and droplet velocities in the range of 1 to 3~m s$^{-1}$.

\subsection{Image analysis}

\begin{figure}[b]
\includegraphics[width=0.5\columnwidth]{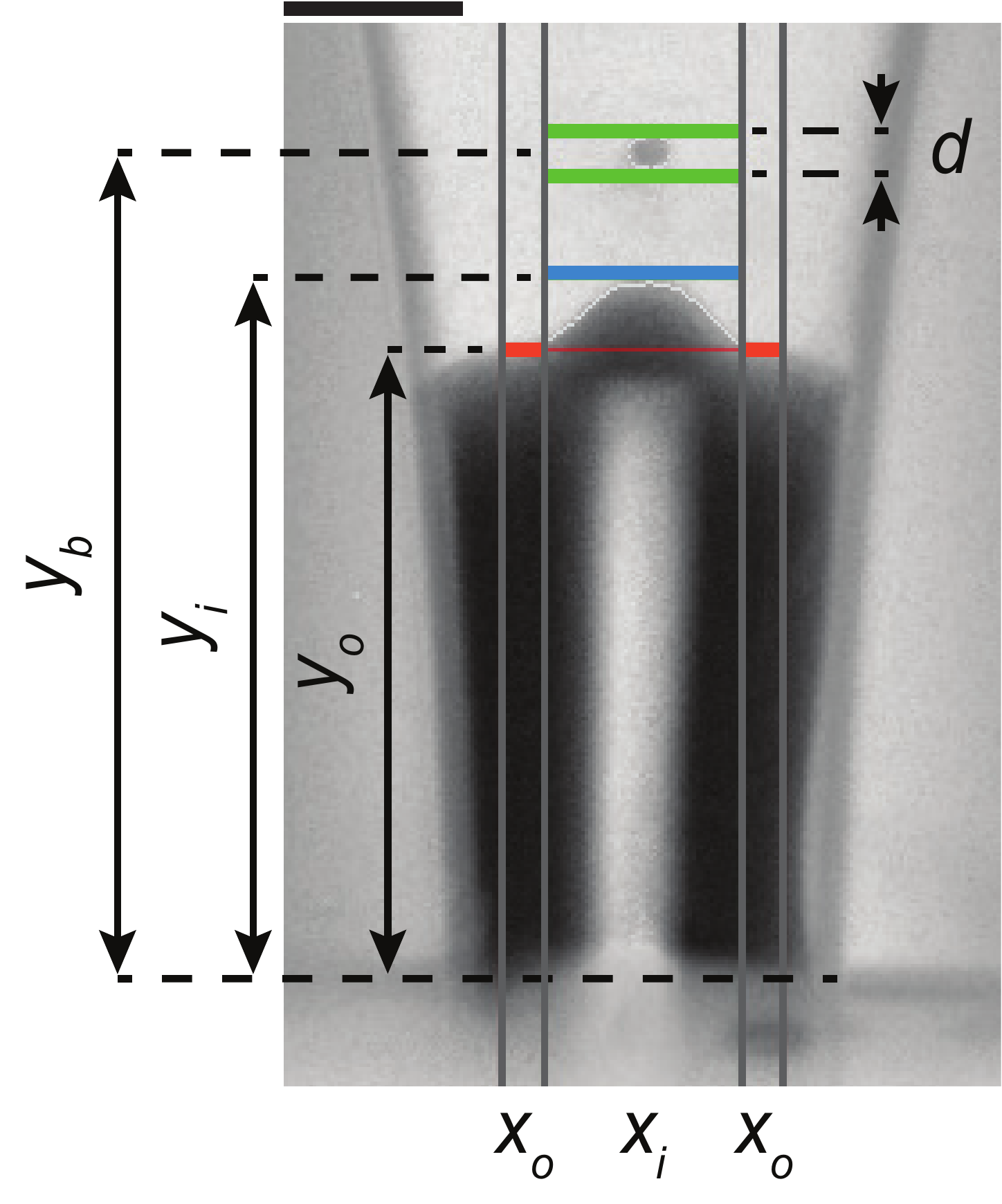}
\caption{Example of an image analysis result, showing the meniscus positions $y_{o}$ of the outer region $x_{o}$ and $y_{i}$ of inner region $x_{i}$, the bubble position $y_{b}$, and the bubble diameter $d$. The scale bar represents 50~\textmu m.}\label{fig:4}
\end{figure}

The motion of the meniscus and that of the bubble were tracked as a function of time. First the contrast in each image was enhanced using ImageJ (http://imagej.nih.gov/ij) by subtracting the original image from the image taken at $t$~=~0~\textmu s, and by adding the inverted result to the original image. Second, the edges of the meniscus and bubble were detected using a script programmed in Python (Python Software Foundation, https://www.python.org/). The script applied a Scikit-Image Canny Edge Detector to each image, extracted the edges of interest, and calculated its positions. The meniscus was separated in an inner and outer region to quantify the meniscus shape deformation, see Fig.~\ref{fig:4}. The inner region was chosen such that it always confined the bubble, and had a width of 0.6~times the nozzle diameter. The outer region was set to a width of 0.9~times the nozzle diameter. The position of the outer region of the meniscus $y_{o}$ was the average position of the detected edge in that region. The position of the inner region of the meniscus $y_{i}$ was the maximum or minimum position of the detected edge in that region depending on whether it had a concave or convex shape, respectively. When a bubble was present, its center position $y_{b}$ and diameter $d$ were determined. The bubble diameter was always measured in axial direction to minimize a potential error in the bubble diameter
due to the refraction of light at the cylindrical walls of the glass nozzle that can deform the image. The time-dependent positions $y_{o}$, $y_{i}$, and $y_{b}$ were filtered to extract the amplitudes and dynamics of the low- and high-frequency components of the meniscus motion.

\subsection{Piezo eigenfrequency characterization through ring-down measurements}

To characterize the resonance behavior of the piezo, the ring-down of the piezo was measured using a piezo sensing technique described in refs.~\cite{dejong2006airentrapment, wijshoff2010inkjet}. The piezo was driven using an electrical pulse. Subsequently, when the driving voltage dropped below 1~V, the piezo contacts were connected automatically to an oscilloscope that recorded the ring-down voltage-oscillations of the piezo giving its characteristic resonance frequency.

\subsection{Boundary integral simulations} \label{BImethod}
To study the bubble pinch-off process and the underlying physical mechanisms in greater detail, boundary integral (BI) simulations were performed~\cite{peters2013highly}. The utilized BI code is axisymmetric, and assumes irrotational, incompressible, and inviscid flow \cite{oguz1993dynamics, power1995boundary, bergmann2009controlled, gekle2009high-speed}. In essence, the unsteady Bernoulli equation:
\begin{equation} \label{Eq:1}
\frac{\partial \phi}{\partial t}=-\frac{1}{2} \left| \nabla \phi \right|^2 - \frac{\Delta p + \kappa \sigma}{\rho},
\end{equation}
is time integrated as described in~\cite{peters2013highly}. In above equation, $\phi$ is the flow potential, $\Delta p$ the pressure variation, $\kappa$ the curvature, and $\sigma$ and $\rho$ the liquid surface tension and density, respectively. Owing to the micron sized meniscus, we neglect gravity. The inviscid assumption is appropriate here as in the experiments it was observed that the meniscus shape deformation is the largest at low viscosity, and decreases as the ink viscosity increases. The numerical setup consisted of a nozzle wall (solid boundary) and a meniscus (free boundary). The flow in the nozzle was driven by applying a stream velocity boundary condition to the nodes at the entrance of the nozzle.

Two methods were used sequentially to describe the contact line dynamics of the meniscus; a fixed contact line, and a moving contact line based on contact angle hysteresis  with a receding contact angle $\theta_{r}$ and an advancing contact angle $\theta_{a}$. Combining these two methods provided a good balance between approximating the experimentally observed meniscus motion and preventing numerical instabilities. These occurred when the distance between the free boundary and solid boundary became too small. At the start of each simulation the contact line was kept pinned. If during this first time period one of the nodes of the meniscus came within a distance from the wall that would cause numerical instability, the meniscus between that node and the contact line was cut off, and a new contact line was created near this node. This intervention did not have a significant effect on the bubble pinch-off phenomena in the simulations. Once the contact angle became larger than $\theta_{a}$, the moving contact line method was initiated. This method keeps the contact line pinned for $\theta_{r} < \theta < \theta_{a}$; moves the contact line to $\theta = \theta_{r}$ if $\theta < \theta_{r}$, and moves the contact line to $\theta = \theta_{a}$ when $\theta > \theta_{a}$. $\theta_{r}$ and $\theta_{a}$ were set to the maximum angle away from 90$^{\circ}$ for which the meniscus motion near the wall remained stable during the simulations, i.e.~$\theta_{r}$~=~72$^{\circ}$ and $\theta_{a}$~=~108$^{\circ}$. At larger angles away from 90$^{\circ}$ numerical instabilities would develop on the meniscus because of the too small distance between the free boundary and the solid boundary, as before. 

\section{Results \& discussion}

\subsection{Meniscus and bubble dynamics}
\begin{figure}[t]
\includegraphics[width=1.0\columnwidth]{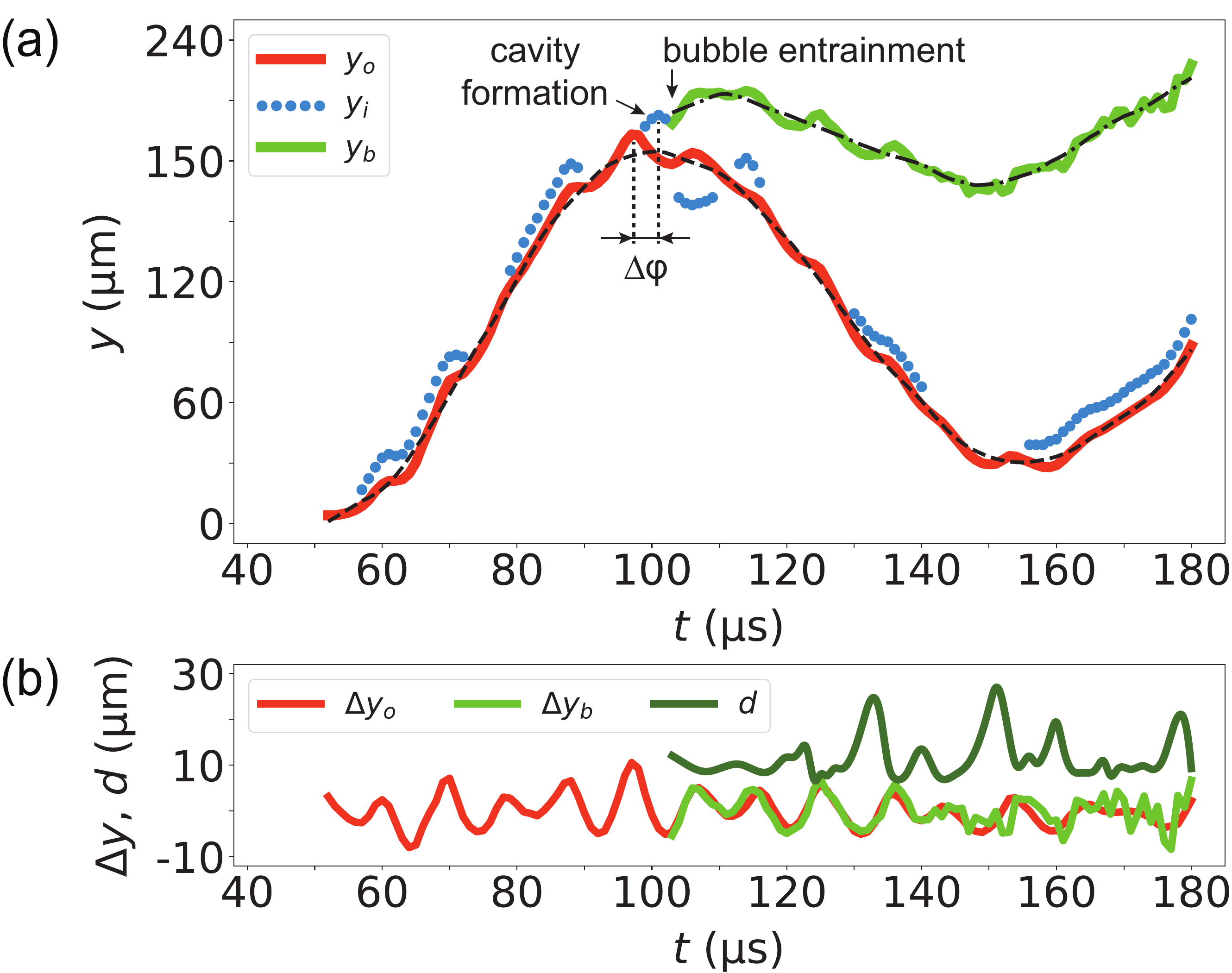}
\caption{(a) Meniscus outer $y_{o}$ and inner $y_{i}$ position, and bubble position $y_{b}$ as function of time after the start of piezo actuation. The black dashed line and the black dash-dotted line show the low-frequency  (low-pass filtered red curve) motion of the meniscus outer region and that of the bubble (low-pass filtered green curve), respectively. (b) High-frequency (high-pass filtered red curve in (a)) movement of the meniscus outer region position $\Delta y_{o}$ and that of the bubble position $\Delta y_{b}$. The dark green curve shows the bubble diameter $d$ as function of time.}\label{fig:5}
\end{figure}

The data shown in Fig.~\ref{fig:1} are now analyzed in more detail. Figure~\ref{fig:5}(a) shows the position of the outer and inner region of the meniscus ($y_{o}$, $y_{i}$) and that of the bubble ($y_{b}$) as a function of time. Note that just before pinch-off a phase difference $\Delta \varphi$ develops between the inner and outer region of the meniscus eventually leading to phase inversion. This is the crucial opposing motion between the central cavity and the outer region of the meniscus that leads to bubble pinch-off, as was observed in Fig.~\ref{fig:1}(b). The process that is responsible for this phase difference is  analyzed in Section~\ref{ssec:deformprocess}. Also note in Fig.~\ref{fig:5}(a) that both the meniscus position curve and the bubble position curve have a high-amplitude low-frequency motion of the order of 10~kHz (100~\textmu s period) with superimposed a low-amplitude high-frequency motion of the order of 100~kHz (10~\textmu s period). The low-frequency motion of the meniscus and bubble are indicated by the dashed curve and by the dash-dotted curve, respectively. 

The high-frequency component in the meniscus motion $\Delta y_{o}$ and that of the bubble motion $\Delta y_{b}$ are plotted in Fig.~\ref{fig:5}(b). In addition, in Fig.~\ref{fig:5}(b) the bubble diameter $d$ is plotted as a function of time. The bubble equilibrium radius was 5~$\pm$~1~\textmu m, which corresponds to a Minnaert eigenfrequency \cite{minnaert1933onmusical} of approximately~600~kHz. As this is much higher than the observed bubble oscillation frequency of 100~kHz, the radial dynamics of the bubble  was considered to oscillate in phase with that of the acoustic pressure waves inside the ink channel~\cite{leighton1994acoustic}. Therefore, the bubble radius directly represents the channel acoustics, i.e.~the maximum in bubble radius corresponds to a minimum pressure, and vice versa. Note in Fig~\ref{fig:5}(b) that the bubble diameter, the bubble position, and the meniscus position all oscillate at a frequency of 105~$\pm$~5~kHz. Also note that the meniscus and bubble are moving inward around the time that the bubble diameter is maximum (pressure minimum), while the meniscus and bubble start moving outward when the bubble diameter is minimum (pressure maximum). Thus, the meniscus and the bubble are driven by the same high-frequency pressure waves, and not by their individual eigenmodes.

\subsection{Acoustic driving by the piezo}
\label{ssec:results-acoustics}

\begin{figure}
\includegraphics[width=1.0\columnwidth]{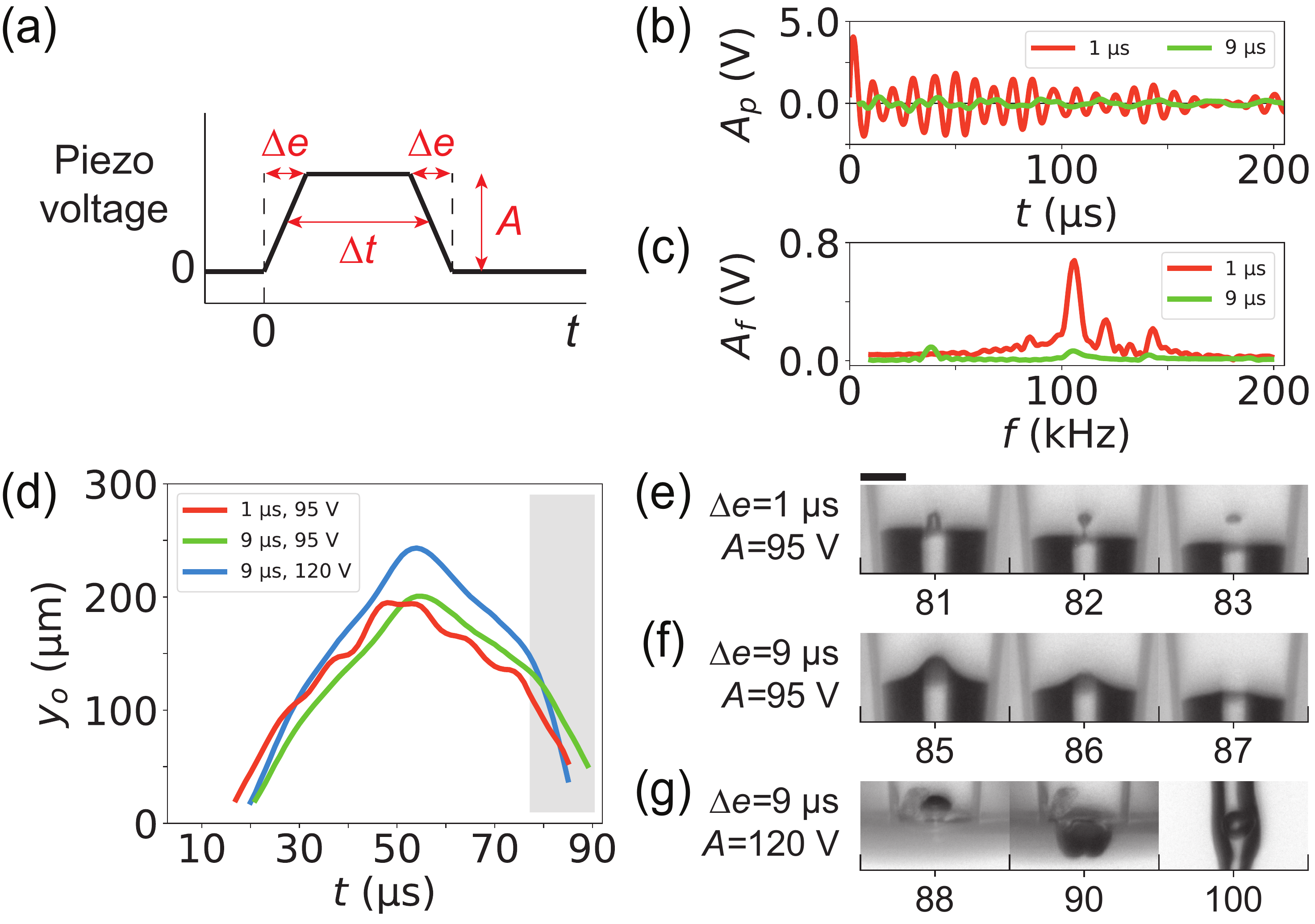}
\caption{(a) Piezo actuation pulse with amplitude $A$, FWHM pulse width $\Delta t$, and rise- and fall time~$\Delta e$. (b) Piezo ring-down measurements with (c) the corresponding Fourier spectra, for a pull-push pulse with $A$~=~10~V, $\Delta t$~=~72~\textmu s, and values of $\Delta e$ as indicated in the legend. (d)~Meniscus motion for three pulses with $\Delta t$~=~72~\textmu s, and with $\Delta e $ and $A$ as given in the legend. Bubble entrainment takes place in the gray-shaded time window. (e, f, g)~Images of the nozzle taken during the time window indicated by the gray-shaded time window in (d), showing whether or not bubble pinch-off took place. The numbers under the images correspond to the time in \textmu s in (d). The scale bar represents 50~\textmu m.}\label{fig:6}
\end{figure}

To determine whether the piezo actuator is the origin of the high-frequency pressure oscillations, the eigenmodes of the piezo were characterized by measuring the ring-down signal of the piezo for an empty ink channel. The piezo was first actuated using a pull-push pulse with an amplitude $A$ of 10~V, a FWHM pulse width $\Delta t$ of 72~\textmu s, and a rise and fall time $\Delta e$ of 1~\textmu s, see Fig.~\ref{fig:6}(a). The ring-down signal and its Fourier spectrum are plotted in Fig.~\ref{fig:6}(b) and \ref{fig:6}(c). Indeed, in the ring-down signal the same 105~kHz high-frequency component was present as in the meniscus motion, bubble motion, and radial dynamics in Fig.~\ref{fig:5}(b). The piezo eigenmode frequencies were calculated from its dimensions (Fig.~\ref{fig:3}) and the speed of sound in piezoceramic ($\sim$4000~m/s) to be 111~$\pm$~12~kHz in longitudinal direction and 3.4~$\pm$~0.1~MHz in radial direction \cite{APCInternational2011piezoelectric, apcpiezocalc}. Thus, the 105~kHz high-frequency component in the piezo ring-down signal originated from the longitudinal resonance mode of the piezo, and the pressure waves produced by this resonance mode drive the high-frequency motion of the meniscus in the nozzle. Indeed, when the high-frequency component is suppressed by using a $\Delta e$ of 9~\textmu s, see Fig.~\ref{fig:6}(b)~and~\ref{fig:6}(c), also the high-frequency motion of the meniscus is suppressed, see Fig.~\ref{fig:6}(d).

Notably, the absence of the 105~kHz high-frequency pressure waves also prevents bubble pinch-off, see Fig.~\ref{fig:6}(f) in comparison to Fig.~\ref{fig:6}(e). In the experiment shown in Fig.~\ref{fig:6}(f) an air cavity was still formed, but it did not pinch-off a bubble. This cavity could be forced to pinch-off in the same way as before, but at a different meniscus position and time, by increasing the amplitude of the piezo driving to 120~V, see Fig.~\ref{fig:6}(g). Thus, suppressing the high-frequency pressure waves effectively increased the threshold for bubble pinch-off from 95~V to 120~V. In other words, the high-frequency pressure waves from the longitudinal resonance mode of the piezo promote bubble pinch-off and through suppression of the 105~kHz waves, the driving amplitude can be increased allowing for stable inkjet printing at a higher droplet velocity.

\subsection{Meniscus shape deformation process}
\label{ssec:deformprocess}

\begin{figure*}
\includegraphics[width=1\textwidth]{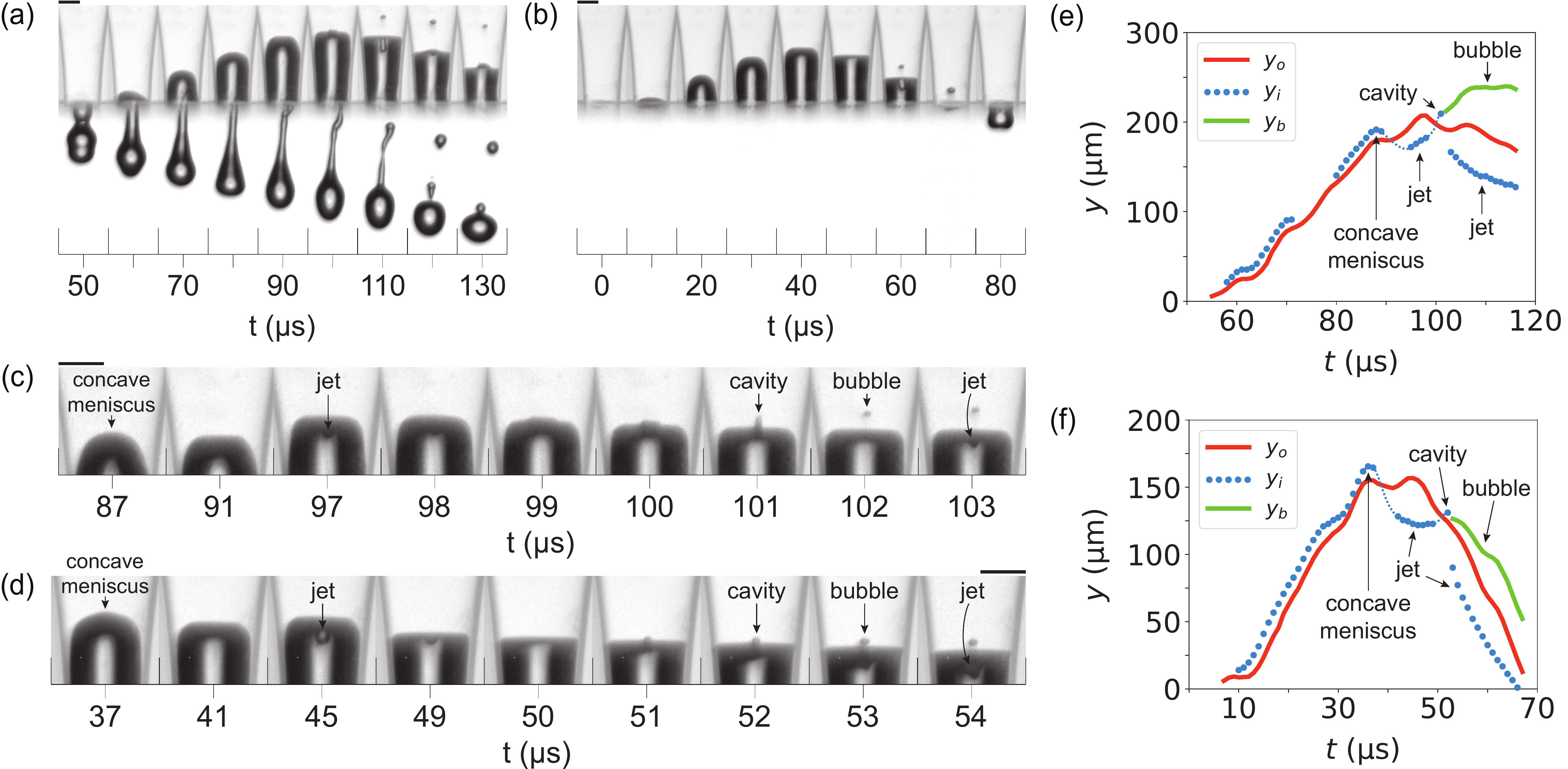}
\caption{Bubble pinch-off for (a) a rectangular push-pull pulse with an amplitude of 160~V and a width of 30~\textmu s, and (b) a rectangular pull-push pulse with an amplitude of 150~V and a width of 30~\textmu s. (c, d) Details of the meniscus shape deformation process prior to, during, and after bubble pinch-off for the push-pull and the pull-push pulse, respectively. (e, f) Meniscus outer region position $y_{o}$, inner region position $y_{i}$, and bubble position $y_{b}$ as function of time for the push-pull and pull-push pulse, respectively. The thin blue dashed line was added to guide the eye in the parts of the jet formation and jet recoil process were the position of the inner region of the meniscus could not be tracked. The scale bars represent 50~\textmu m.}\label{fig:7}
\end{figure*} 

Now that the driving mechanisms of the meniscus are identified, next, the process responsible for the development of the phase difference between the inner and outer region of meniscus, eventually leading to phase inversion and bubble pinch-off, can be identified. The phase difference $\Delta \varphi$ (Fig.~\ref{fig:5}(a)) develops through the meniscus shape deformation process that can be observed in Fig~\ref{fig:7}: it develops by jet formation at a concave meniscus. The universality of the meniscus shape deformation process prior to bubble pinch-off is demonstrated in Fig.~\ref{fig:7} by its presence in two bubble pinch-off experiments with entirely different driving conditions, namely, with piezo driving pulses with opposite polarity. In the first experiment the piezo was actuated using a rectangular push-pull pulse ($A$=~160~V, $\Delta t$~=~30~\textmu s). A bubble was entrained after droplet formation, and it remained inside the nozzle, see Fig.~\ref{fig:7}(a). In the second experiment the piezo was actuated using a rectangular pull-push pulse ($A$~=~150~V, $\Delta t$~=~30~\textmu s). In contrast to the first experiment, a bubble was entrained before droplet formation, and it was ejected with the droplet shortly after entrainment, see Fig.~\ref{fig:7}(b). Despite the large differences between the two experiments, the image sequences and graphs in Fig.~\ref{fig:7}(c-f) show that the meniscus shape deformation process is qualitatively the same for the two experiments. Initially, upon retraction, the meniscus has a concave shape. Then, during the advancing of the meniscus, a small liquid jet is formed in outward direction. Later, this jet recoils back inward, while the outer region of the meniscus is forced to move outward again, in the opposite direction of the movement of the jet. Similar to the experiment in Fig.~\ref{fig:1}, the opposing motion of the outer and inner region of the meniscus leads to the formation and closure of a cavity, and thereby to the pinch-off of a bubble. Thus, phase inversion between the inner and outer region of the meniscus is a consequence of jet formation at the central part of the concave meniscus.

\subsection{Jet formation mechanism}
From literature it is known that when a pressure wave propels a concave-shaped meniscus forward, a jet forms due to geometrical focusing of the flow at the meniscus due to an inhomogeneous pressure gradient field along the meniscus~\cite{peters2013highly,Gordillo2020}. The pressure gradient and resulting velocity are larger at the center of a concave meniscus than at its edge, see also ref.~\cite{antkowiak2007short-term}. Thus, in the inkjet nozzle, first, the inward motion of ink results in a concave shaped meniscus, then a first outward acceleration creates a phase difference between the inner and outer region of the meniscus by the formation of a central outward-moving liquid jet, and, finally, a well-timed second outward acceleration enhances this phase difference by the formation of a toroidal outward-moving liquid jet. The central liquid jet recoils inward and forms an air cavity that is enclosed by the toroidal outward-moving liquid jet, and as a consequence, a bubble pinches off.

\begin{figure*}[t]
\includegraphics[width=.91\textwidth]{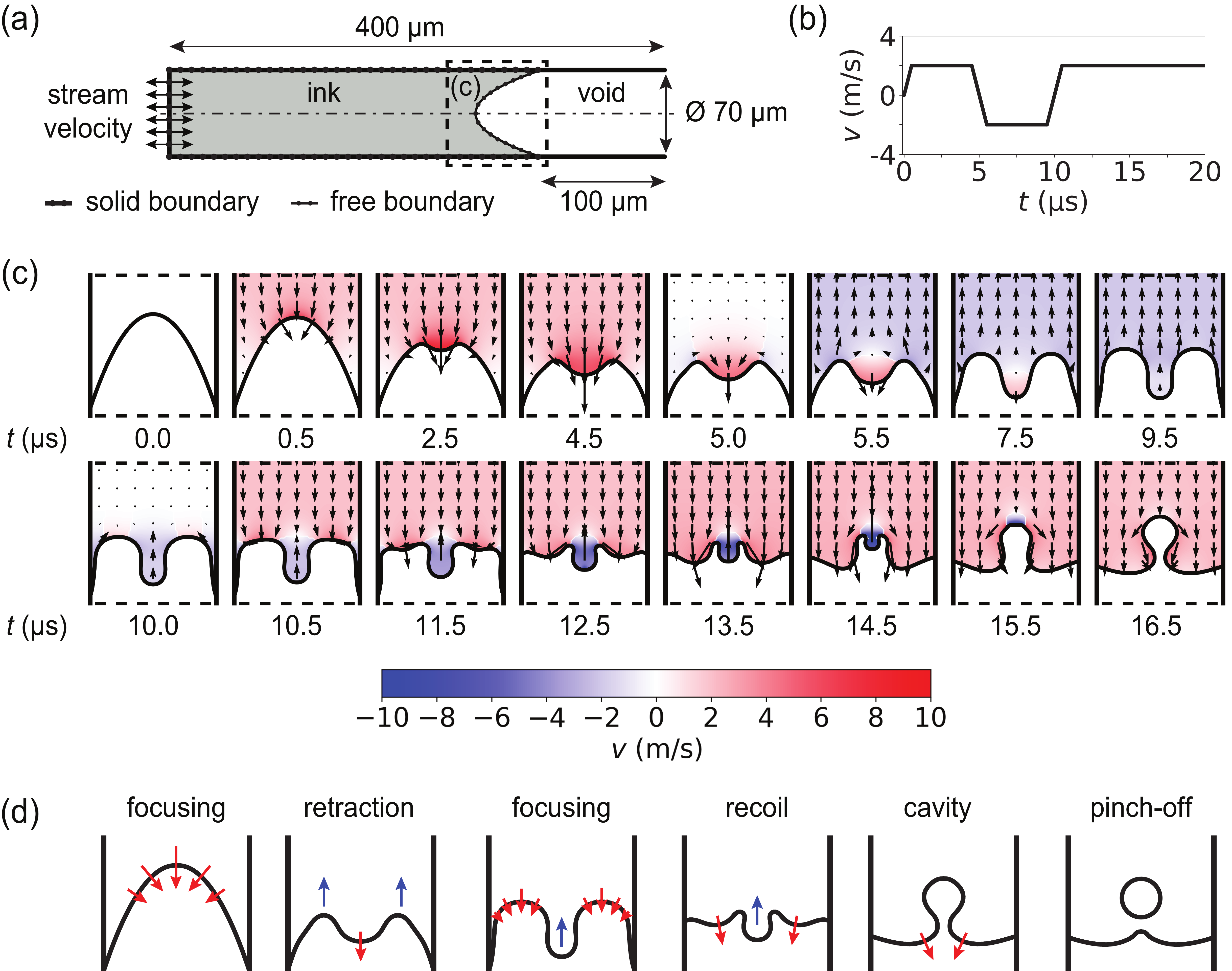}
\caption{(a) Numerical setup for the boundary integral (BI) simulation where the left boundary is subject to a stream velocity boundary condition. The initial meniscus shape is a parabola with a depth of 0.75~times the nozzle diameter. (b) Stream velocity boundary condition $v$ for the numerical setup as function of time, mimicking a 100 kHz pressure oscillation, followed by an outward directed flow. (c) Meniscus shape deformation process prior to bubble pinch-off, simulated using the BI method. (d) Schematic summary of the main steps in (c).}\label{fig:8}
\end{figure*}

To further demonstrate the details of the proposed pinch-off mechanism, numerical simulations were performed using the boundary integral (BI) method. The results are shown in Fig.~\ref{fig:8}. The geometry of the numerical setup in Fig.~\ref{fig:8}(a), and the stream velocity boundary condition $v(t)$ in Fig.~\ref{fig:8}(b), were chosen such that they follow the experimental conditions, i.e., the nozzle diameter and the 110~kHz oscillations that were identified to drive bubble pinch off. The initial meniscus shape (Fig.~\ref{fig:8}(c)) together with the imposed amplitude of the 110~kHz velocity boundary condition (Fig.~\ref{fig:8}(b) were varied and, as in the experiments, only for highly specific combinations of the two, the meniscus shape deformation process was developing toward bubble pinch-off. A simulation result is shown in Fig.~\ref{fig:8}(c). The figure reveals the amplitude and the direction of the ink velocity, and demonstrates how the velocity field inhomogeneity and the focusing of the flow at the concave part of the meniscus drive jet formation. Moreover, Fig.~\ref{fig:8}(c) highlights the opposing motion between the central jet and the toroidal jet, and shows in detail how this opposing motion leads to the formation of a cavity that closes and pinches off a bubble. The main steps in this process, which were discussed in detail before, are now schematically summarized in Fig.~\ref{fig:8}(d): a central jet forms at the concave meniscus during the first outward acceleration. Then, a toroidal jet forms at the concave meniscus around the central jet during the second outward acceleration. The recoiling central jet forms a cavity, and the progressing toroidal jet encloses this cavity, with bubble pinch-off as a result.

Note that even though the boundary integral simulations are incompressible, we observe the same meniscus deformation process as in the experiments suggesting that the meniscus shape deformation process responsible for bubble pinch off does not necessarily require acoustics, but that a certain unbalance between the capillary and inertial time scales is sufficient. Indeed, the capillary time scale ($\tau_c = \sqrt{\rho R^3/\sigma}$, with $R$ the nozzle radius) for the present inkjet nozzle is approx. 24~$\mu$s whereas the inertial timescale is of the order of 1~$\mu$s, i.e., set by the rise and fall time of $v(t)$ in Fig.~\ref{fig:8}(b). Furthermore, from the input velocity boundary condition $v(t)$ and Eq.~\ref{Eq:1} the dynamic driving pressure in the BI simulations can be calculated, as follows: $\Delta p = \Delta p_a + \Delta p_i = - \rho \frac{\partial \phi}{\partial t} - \tfrac{1}{2}\rho U^2$, with $\Delta p_a$ and $\Delta p_i$ the pressure contributions from acceleration and inertia, respectively. The magnitude of $\Delta p_i$ is directly estimated from $v(t)$ as 8~kPa. The magnitude of $\Delta p_a$ can be estimated as $\rho U L/\Delta t \approx$ 1.4 to 12~bar, with $U$ the velocity variation (4~m/s) over time $\Delta t$ (1~$\mu$s, see Fig.~\ref{fig:8}(b)) and $L$ a length scale between the nozzle radius (35~$\mu$m) and the fluid-filled domain (300~$\mu$m, see Fig.~\ref{fig:8}{a}). Note that since the BI simulations are incompressible, these dynamic pressure fluctuations are to be compared with acoustic pressure fluctuations in the experiment. The order of magnitude of the dynamic driving pressure amplitude is in line with values reported for the acoustic driving pressure in inkjet printing~\cite{wijshoff2010inkjet,Fraters2019Nucleation} which once more shows that acoustic wave propagation is not required for the observed bubble pinch-off phenomenon and that it is mainly a flow-dominated process. 

\subsection{The bubble pinch-off window}
\begin{figure}[t]
	\includegraphics[width=1\columnwidth]{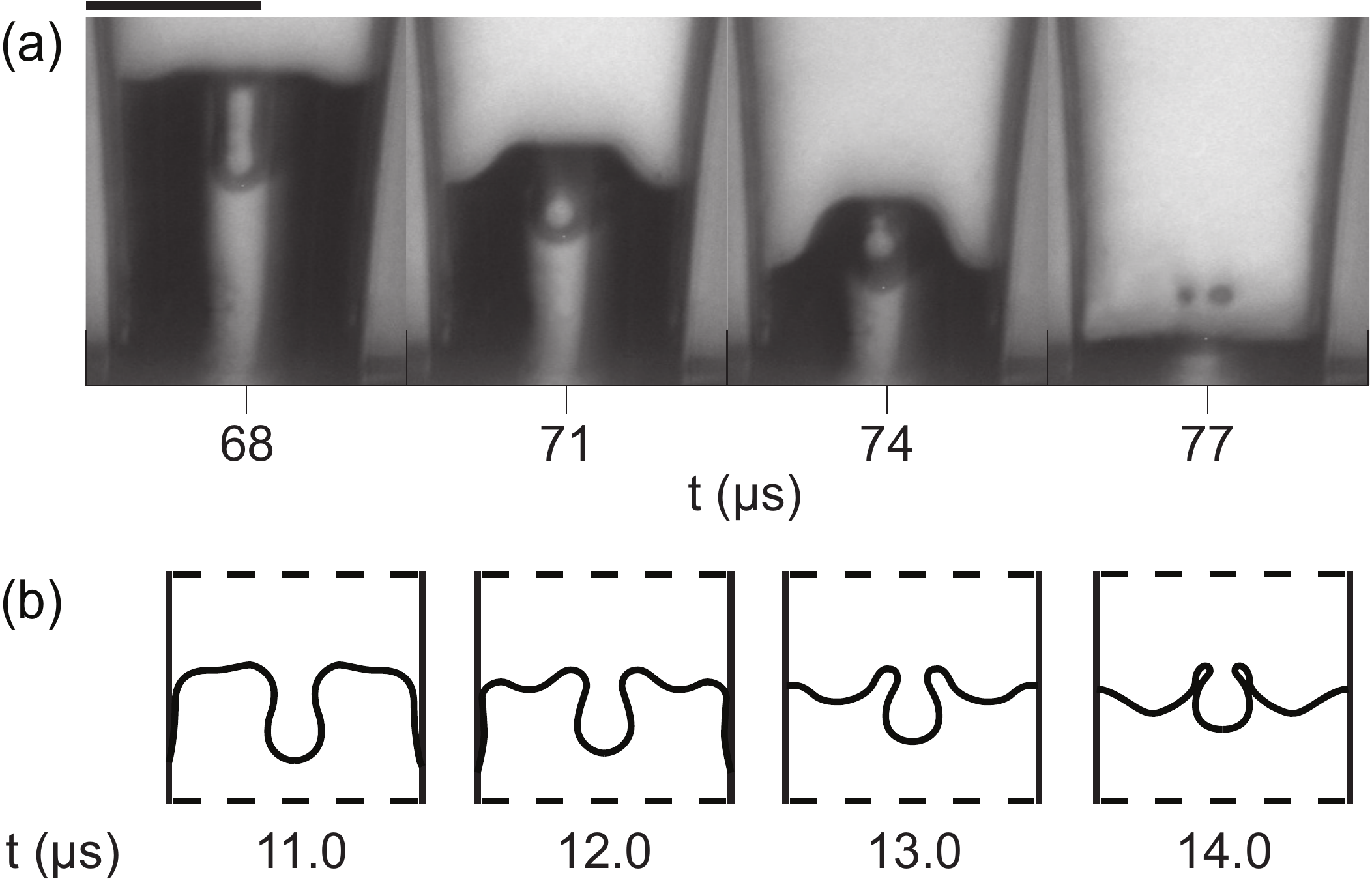}
	\caption{ (a) Toroidal bubble pinch-off as experimentally recorded for a rectangular pull-push pulse with an amplitude of 92~V and a width of 53~\textmu s. The scale bar represents 50~\textmu m. (b) Simulation using the BI method of the meniscus shape deformation process eventually leading to toroidal bubble pinch-off.}\label{fig:9}
\end{figure}

Using the acquired knowledge on the underlying physics of the bubble pinch-off phenomenon studied in this work, we now qualitatively explain the observed parameter windows of bubble pinch-off in Fig.~\ref{fig:2}. In Fig.~\ref{fig:2}(b) the pulse width $\Delta t$~was fixed and the amplitude $A$~was varied. At an amplitude of 130~V and lower, the velocity difference between the recoiling central jet and the progressing toroidal jet was not high enough to form a sufficiently deep cavity at the right moment in time and to enclose this cavity. At an amplitude of 160~V the central jet had such a length and inertia that it was too slow to recoil before the toroidal jet reached the central axis. As a result, the toroidal jet enclosed the base of the central jet, which in multiple experiments and simulations has been observed to result in the formation of a toroidal bubble such as shown experimentally in Fig.~\ref{fig:9}(a) and from a BI simulation in Fig.~\ref{fig:9}(b).

In Fig.~\ref{fig:2}(c) the amplitude $A$~was fixed, and the pulse width $\Delta t$~was varied. In other words, the control parameter in these experiments was the timing of the outward acceleration of the meniscus by the falling edge of the piezo driving pulse. In the experiments shown in Fig.~\ref{fig:2}(c) the central jet had already been formed before the falling edge of the pulse. At the different times of meniscus acceleration, the meniscus shape was different, and thus the toroidal jet formation process was different. At $\Delta t$~=~69~\textmu s the acceleration was too early, i.e.~the central jet was not able to develop sufficient opposing motion with respect to the toroidal jet because of its early formation. At $\Delta t$~=~76~\textmu s the acceleration was too late, i.e.~the meniscus was propelled outwards while the cavity was already present, thus, the central cavity was propelled outward faster than the outer region of the meniscus.

Despite the acquired knowledge on the underlying physics of the bubble pinch-off mechanism, it remains difficult to predict where exactly in the piezo driving parameter space bubble pinch-off will occur, as is also the case for bubble pinch-off after drop impact on a liquid pool~\cite{Oguz1990,Sleutel2020}. The two main reasons are the sensitivity of the mechanism to the operating conditions and the unavailability of information about the exact printhead configuration and its detailed acoustic properties. As a practical guideline, however, bubble pinch-off can be suppressed with relative ease by suppressing the high-frequency component in the acoustics through the edge duration of the piezo driving pulse. Another simple method, which was not studied here, is to damp out the meniscus shape deformations by increasing the ink viscosity. However, this requires higher driving amplitudes to produce droplets at equal velocity and reduces the universal applicability of the technique. The role of ink viscosity on meniscus deformations due to flow focusing in an inkjet nozzle will be part of future work.

\section{Conclusions}
An acoustically driven meniscus in a piezo inkjet nozzle can pinch-off a bubble under specific driving conditions. Pinch-off is the result of the closure of a central air cavity in the meniscus that forms due to opposing motion between a central region and an outer region of the meniscus. The opposing motion between the central region and outer region of the meniscus is the result of jet formation at the concave meniscus. Jet formation is driven by flow focusing, i.e., due to the inhomogeneous pressure gradient field along the meniscus, as was confirmed by boundary integral simulations. The process that is responsible for the bubble pinch-off can be summarized as follows: the meniscus gains a concave shape due to inward motion. Subsequently, a first outward acceleration produces a central jet at the concave meniscus. A well-timed second outward acceleration produces a toroidal jet at the concave meniscus around the central jet. The recoiling central jet forms a central air cavity while the progressing toroidal jet encloses this air cavity. Eventually, this leads to pinch-off of an air bubble. These results gain fundamental understanding of the stability of an acoustically driven meniscus in an inkjet printhead and thereby provide ways to increase the stability of inkjet printing. The results from the incompressible boundary integral simulations suggest that bubble pinch off requires a certain unbalance between the capillary and inertial time scales and that it is mainly flow-driven and does thereby not necessarily require acoustics. Future work will focus on further elucidating the role of acoustics and liquid viscosity on bubble pinch off and entrainment.

\begin{acknowledgments}
This work is part of the research program "High Tech Systems and Materials" (HTSM) with project number 12802, and part of the Industrial Partnership Program number i43, of the Dutch Technology Foundation (STW) and the Foundation for Fundamental Research on Matter (FOM), which are part of the Netherlands Organisation for Scientific Research (NWO). The research was co-financed by  Canon Production Printing Netherlands B.V., University of Twente, and Eindhoven University of Technology.
\end{acknowledgments}

%

\end{document}